\documentclass{WileyMSP-template}

\usepackage{cite} 

\usepackage{ragged2e}
\usepackage{etoolbox}
\usepackage{makecell} 
\usepackage{array}

\usepackage{booktabs}
\usepackage{multirow}
\usepackage{array}
\usepackage{siunitx}
\usepackage{caption}

\begin{document}
\setlength{\parfillskip}{0pt plus 1fil} 
\setlength{\emergencystretch}{3em}

\pagestyle{fancy}
\rhead{\includegraphics[width=2.5cm]{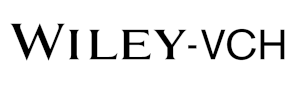}}

\title{Giant Tunneling Magnetoresistance in Graphene/$h$-BN Based van der Waals Magnetic Tunnel Junctions via 3$d$ Transition Metal Intercalation}

\maketitle


\author{Zhi Yan*}
\author{Jianhua Xiao}
\author{Xujin Zhang}
\author{Cheng Fang}
\author{Xiaohong Xu*}




\begin{affiliations}
Zhi Yan, Jianhua Xiao, Xujing Zhang, Cheng Fang, Xiaohong Xu\\
School of Chemistry and Materials Science $\&$ Key Laboratory of Magnetic Molecules and Magnetic Information Materials of Ministry of Education, Shanxi Normal University, Taiyuan 030031, China\\
Zhi Yan, Xiaohong Xu\\
School of Chemistry and Materials Science $\&$ Key Laboratory of Magnetic Molecules and Magnetic Information Materials of Ministry of Education, Shanxi Normal University, Taiyuan 030031, China\\
Research Institute of Materials Science $\&$ Shanxi Key Laboratory of Advanced Magnetic Materials and Devices, Shanxi Normal University, Taiyuan 030031, China\\
Email Address: yanzhi@sxnu.edu.cn; xuxh@sxnu.edu.cn

\end{affiliations}



\begin{justify}

\begin{abstract}
Atomic intercalation offers a powerful route for engineering two-dimensional (2D) materials by precisely tuning interlayer electronic coupling and spin configurations. Here, we propose a generic strategy for the construction of fully 2D magnetic tunnel junctions (MTJs) based on transition metal-intercalated graphene electrodes with $h$-BN barrier layer. First-principles calculations reveal that intercalation not only stabilizes uniform atomic dispersion via steric hindrance but also induces robust ferromagnetism in graphene. Manganese- and vanadium-intercalated systems (Mn-Gr and V-Gr) exhibit exceptional spintronic performance, with tunneling magnetoresistance (TMR) showing a pronounced odd-even oscillation as a function of barrier thickness. A giant TMR of $4.35 \times 10^8\,\%$ is achieved in the Mn-Gr system with a monolayer barrier $h$ -BN ($n=1$), while V-Gr reaches a maximum TMR of $1.86 \times 10^5\,\%$ for a trilayer barrier ($n=3$). 
Moreover, biaxial strain further enhances the TMR to $10^9\,\%$ and $10^7\,\%$ in Mn-Gr and V-Gr systems, respectively. The devices also exhibit perfect spin filtering and pronounced negative differential resistance, offering new opportunities for high-performance spintronic and memory applications based on 2D van der Waals heterostructures.

\end{abstract}


\section{Introduction}

Magnetic tunnel junctions (MTJs) are central components in spintronic technologies, widely employed in read heads, magnetic sensors, and non-volatile memories due to their high scalability and endurance~\cite{ref1,ref2,ref3,ref4}. A key figure of merit for MTJ performance is the tunneling magnetoresistance (TMR) ratio, which directly impacts device functionality and efficiency~\cite{ref5,ref6,ref7}. However, conventional MTJ architectures face two major challenges: the need for increased storage density and aggressive device miniaturization, both of which exacerbate the impact of interfacial lattice mismatch and defect-induced spin scattering, thereby significantly degrading transport performance~\cite{ref8,ref9,ref10,ref11,ref12,ref13}.

Two-dimensional (2D) van der Waals (vdW) materials offer an appealing route to overcome these limitations owing to their atomically flat surfaces and weak interlayer interactions~\cite{ref14,ref15,ref16,ref17,ref18}. These intrinsic properties facilitate the fabrication of high-quality heterostructures without the constraints of lattice matching, enabling atomically sharp interfaces crucial for spintronic applications. Nevertheless, the lack of robust intrinsic magnetism in most synthesized 2D materials remains a fundamental bottleneck. Existing 2D magnetic materials are scarce and typically suffer from low Curie temperatures and poor ambient stability, which severely restrict their practical deployment in high-performance spintronic devices~\cite{ref19,ref20}.

Atomic intercalation has emerged as a powerful strategy to engineer both the electronic and magnetic properties of 2D systems. Unlike conventional chemical doping, intercalation can induce localized magnetic moments or long-range magnetic ordering while minimizing lattice distortions. This is enabled by the unique structural characteristics of 2D materials: strong in-plane covalent bonding coexisting with weak out-of-plane vdW interactions, which collectively create interlayer galleries that can host foreign atoms without disrupting the host lattice~\cite{ref21,ref22,ref23,ref24}. Strategic selection and spatial distribution of intercalants allow precise control over magnetic anisotropy, Curie temperature, and electronic phases. A prominent example is lithium-intercalated MoS$_2$ (Li$_x$MoS$_2$), where sufficient charge injection drives a phase transition from the semiconducting 2H phase to the metallic 1T structure~\cite{ref25}. Recent theoretical studies have shown that transition metal intercalation into bilayer transition metal dichalcogenides significantly modifies their electronic states and magnetic ground states, offering a blueprint for next-generation spintronic platforms~\cite{ref26}. Experimentally, He \emph{et al.} synthesized Fe-intercalated TaS$_2$ (Fe$_{1/3}$TaS$_2$) via flux growth and observed gate-tunable ferromagnetism, validating the feasibility of intercalation-induced magnetic order in 2D systems~\cite{ref27}.

In this work, we theoretically investigate the spin-dependent electronic transport properties of vdW MTJs based on 3$d$ transition metal-intercalated graphene by employing the nonequilibrium Green’s function combined with density functional theory. Our results reveal that intercalation not only induces robust ferromagnetism in graphene but also suppresses metal atom aggregation via steric hindrance. The transport properties are found to be highly sensitive to the thickness of the $h$-BN barrier, with TMR exhibiting pronounced odd-even oscillations in Mn- and V-intercalated graphene junctions (Mn-Gr and V-Gr). A giant TMR of $4.35 \times 10^8\%$ is achieved in the Mn-Gr system with a monolayer barrier, while the V-Gr system reaches a maximum TMR of $1.86 \times 10^5\%$ with a trilayer barrier. Furthermore, the application of $\pm3\%$ biaxial tensile strain dramatically enhances TMR, elevating it to the $10^9\%$ and $10^7\%$ levels in Mn-Gr and V-Gr systems, respectively. The observed perfect spin filtering and negative differential resistance (NDR) effects highlight the potential of atomic intercalation engineering as a viable route toward scalable, high-performance, and fully 2D spintronic devices.

\section{Results and Discussion}
\subsection{Design of vdW MTJs}
Since the isolation of graphene in 2004~\cite{ref28}, extensive efforts have been devoted to realizing room-temperature ferromagnetism for potential spintronic applications. However, the lack of intrinsic magnetic moments in graphene remains a fundamental challenge. Conventional approaches such as chemical doping~\cite{ref29} and surface adsorption~\cite{ref30} often suffer from magnetic atom clustering and antiferromagnetic coupling, hindering long-range magnetic ordering. In contrast, atomic intercalation into bilayer graphene offers a promising route: the confined, periodic interlayer space suppresses atomic clustering and favors ordered occupation of high-symmetry sites, as shown in Fig.~\ref{fig1}.
Building on the intercalation strategy described above, we systematically explored six high-symmetry intercalation sites (see Fig.~S1) for 3\emph{d} transition metals (Sc–Zn). Configurations exhibiting significant lattice distortions or spin-degenerate states were excluded. The energetically most favorable and structurally stable configurations were identified (see Fig.~S2), with Mn- and V-intercalated bilayer graphene (Mn–Gr and V–Gr, respectively) emerging as promising candidates due to their stability and robust ferromagnetism.
As shown in Figs.~\ref{fig2}(a) and \ref{fig2}(b), the top and side views of the relaxed structures, along with their spin-resolved band structures, reveal clear spin splitting. While pristine graphene exhibits spin-degenerate bands, intercalation of transition metal atoms lifts this degeneracy, resulting in a distinct separation between spin-up and spin-down channels. This spin splitting originates from the unpaired 3\emph{d} electrons of the intercalated atoms, which retain finite magnetic moments and impart ferromagnetic characteristics to the graphene host.
To further assess the aggregation tendency of intercalated Mn and V atoms, we constructed initial cluster-like configurations with 2, 3, 4, and 5 atoms inserted into bilayer graphene, as illustrated in Fig.~S3. After structural relaxation, the interatomic distances increased significantly and the atoms became uniformly distributed, indicating a strong suppression of clustering. These results highlight the effectiveness of interlayer spatial confinement in preventing three-dimensional aggregation via steric hindrance, thereby stabilizing well-dispersed magnetic configurations at the atomic scale.
The above results confirm that 3$d$ transition metal intercalated graphene can serve as a viable magnetic electrode in MTJs.

Hexagonal boron nitride ($h$-BN), sharing a similar two-dimensional layered structure with graphene~\cite{ref31,ref32}, possesses a closely matched lattice constant (2.50~\AA\ for $h$-BN vs. 2.46~\AA\ for graphene)~\cite{ref33,ref34}. This structural compatibility minimizes defect formation and enhances interfacial stability when forming van der Waals heterostructures, suggesting that $h$-BN is a promising candidate for the tunneling barrier in MTJs.
Based on the above insights, we constructed two types of fully vdW MTJs using the simplest graphene/\textit{h}-BN heterostructure: Mn-Gr/\textit{n}$\cdot$\textit{h}-BN/Mn-Gr and V-Gr/\textit{n}$\cdot$\textit{h}-BN/V-Gr (\textit{n} = 1-4) vdW MTJs, as illustrated in Figs.~\ref{fig2}(e) and 2(f). Each device consists of symmetric 3\textit{d} transition metal–intercalated graphene electrodes and a tunable barrier layer composed of \textit{n} layers of \textit{h}-BN, with periodic boundary conditions applied in the \textit{xy}-plane. As shown in Fig.~\ref{fig2}(c), we examined three representative high-symmetry stacking configurations at the graphene/\textit{h}-BN interface: (i) N-hollow, where nitrogen atoms align with the hollow sites of the graphene lattice; (ii) B-hollow, with boron atoms positioned at hexagon centers; and (iii) BN-C, corresponding to AA-type stacking between \textit{h}-BN and graphene. Among them, the N-hollow configuration exhibits the lowest energy. In addition, four bilayer \textit{h}-BN stacking geometries were considered, and Fig.~\ref{fig2}(d) confirms that the N-B, B-hollow arrangement is energetically most favorable. These results establish the structural foundation of two fully vdW MTJs based on minimal graphene/\textit{h}-BN heterostructures.

\subsection{TMR effects and biaxial strain in equilibrium state}
MTJs exhibit two distinct spin configurations determined by the relative alignment of magnetizations in the electrodes: parallel (P) and antiparallel (AP). In our simulations, the left electrode and its adjacent buffer layer were fixed, while the right electrode served as the free layer. The tunnel magnetoresistance (TMR), defined as $\mathrm{TMR} = (G_{\mathrm{P}} - G_{\mathrm{AP}})/G_{\mathrm{AP}}$, quantifies the spin-dependent transport performance. We systematically evaluated the equilibrium TMR characteristics of Mn-intercalated (Mn-Gr/\textit{n}·\textit{h}-BN/Mn-Gr) and V-intercalated (V-Gr/\textit{n}·\textit{h}-BN/V-Gr) vdW MTJs, as summarized in Table~\ref{tab1}. Notably, Mn-based devices exhibit significantly higher TMR ratios compared to their V-based counterparts. This discrepancy originates from the distinct electronic configurations of the 3\textit{d} transition metals: this enhancement originates from the half-filled $3d^5$ configuration of Mn atoms, which leads to a strong localized magnetic moment and significant spin splitting of graphene’s Dirac cone due to charge transfer and hybridization effects.

Remarkably, both Mn-Gr/\textit{n}·\textit{h}-BN/Mn-Gr and V-Gr/\textit{n}·\textit{h}-BN/V-Gr MTJs exhibit pronounced odd–even oscillations in TMR as a function of barrier thickness. The Mn-based MTJs achieve an exceptional TMR ratio of $4.35 \times 10^8\%$ at the monolayer barrier ($n = 1$), with the TMR magnitude decreasing in an odd–even modulated fashion as the number of \textit{h}-BN layers increases. In contrast, the V-based MTJs exhibit nearly perfect spin filtering in the parallel configuration, reaching a peak TMR of $1.86 \times 10^5\%$ at $n = 3$ layers. These results underscore the outstanding equilibrium-state spintronic performance of atomically engineered vdW MTJs, reinforcing their promise for high-efficiency non-volatile memory technologies.

The characteristic TMR effect is distinctly observed in the projected density of states (PDOS) analysis. The electronic structure of the central region was systematically examined through real-space PDOS mapping as a function of the Fermi energy (\textit{E$_F$}) and vertical transport distance (\textit{z}), as shown in Fig.~3. The horizontal axis represents the spatial extent of the central region, where the dark blue region near \textit{z} = 10 \AA\ corresponds to the tunneling barrier induced by the insulating \textit{h}-BN layer. This spatial PDOS distribution confirms that \textit{h}-BN provides an adequate tunneling barrier height, with electron transport being governed primarily by quantum tunneling mechanisms.
Figures ~\ref{fig3}(a)-3(d) and 3(e)-3(h) depict the PDOS distributions for Mn-Gr/\textit{n}·\textit{h}-BN/Mn-Gr and V-Gr/\textit{n}·\textit{h}-BN/V-Gr vdW MTJs, respectively. Taking Mn-Gr/\textit{h}-BN/Mn-Gr MTJs (with \textit{n} = 1) as a representative case (see Fig.~\ref{fig3}a), we observe that spin-up channels are suppressed at the Fermi level (\textit{E$_F$}) in the P configuration, while spin-down channels dominate on both Mn-Gr electrodes at \textit{E$_F$}, resulting in a low-resistance state for electron transport. When the system is switched to the AP configuration, spin-up (spin-down) electrons propagate from the left majority (minority) states through the \textit{h}-BN barrier to the right minority (majority) states at \textit{E$_F$}. This interfacial mismatch in electronic state parity between the left and right Mn-Gr layers leads to the formation of a high-resistance state. A similar mechanism governs electron transport in the V-Gr/\textit{n}·\textit{h}-BN/V-Gr MTJs, demonstrating the universality of this quantum transport phenomenon and proving the characteristic TMR effect.

Momentum-resolved quantum transport analyses were performed by calculating the $k_{\parallel}$-resolved transmission coefficients at the E$_F$ within the two-dimensional Brillouin zone (2D-BZ) for both Mn-Gr/\textit{n}·\textit{h}-BN/Mn-Gr and V-Gr/\textit{n}·\textit{h}-BN/V-Gr vdW MTJs, with the transport direction along the $z$-axis. As shown in Figs. ~\ref{fig4}(a)-4(d) and 4(e)-4(h), a pronounced disparity in transmission 'hot spots' was observed between the P and AP configurations in Mn-Gr-based MTJs, where the AP-state 'hot spots' nearly vanish—an effect directly correlated with the observed giant TMR. In V-Gr-based MTJs, spin-down channels exhibited severely suppressed transmission probabilities in the P-state, while spin-up channels maintained high transmission coefficients, demonstrating near-ideal spin filtering effect. Furthermore, the monotonic attenuation of the 'hot spots' density with increasing \textit{h}-BN layer thickness was attributed to reduced tunneling probabilities due to the enhanced barrier thickness. The distribution of transmission coefficients in the 2D-BZ provides additional evidence supporting the presence of a giant TMR ratio and perfect spin-filtering effect in these MTJs in the equilibrium state.

Strain-engineered quantum modulation has been shown to effectively optimize magnetoelectric transport properties in vdW tunnel junctions through the application of in-plane biaxial strain \cite{ref35, ref36, ref37, ref38}. We also systematically investigate the influence of biaxial strain ($\varepsilon = -3\%$ to +3\%, with 1\% increments) on Mn/V-intercalated graphene-based MTJs with monolayer \textit{h}-BN barriers (\textit{n} = 1). As depicted in Fig.~\ref{fig5}, strain modulation enables precise control over TMR and spin injection efficiency (SIE). For Mn-Gr/\textit{h}-BN/Mn-Gr MTJs, TMR magnitudes exhibited a decrease under compressive strain but increased under tensile strain, reaching up to $10^9\%$. In contrast, SIE in the P configuration showed an inverse correlation with TMR trends, while the AP configuration SIE exhibited minimal strain sensitivity. For V-Gr/\textit{h}-BN/V-Gr MTJs, TMR remained stable within the $\varepsilon = -3\%$ to +1\% strain range but surged to $10^7\%$ at +3\% tensile strain. Notably, SIE in the P configuration maintained 100\% spin polarization across all strain conditions, confirming strain-robust ideal spin filtering. In contrast, the AP configuration SIE demonstrated oscillatory behavior, peaking at 91\% under +1\% tensile strain. These results establish strain engineering as a promising strategy for tailoring quantum transport in vdW MTJs, offering new opportunities for optimizing spintronic device performance.

\subsection{Spin-dependent transport properties of MTJs in the nonequilibrium state}
To further elucidate the nonequilibrium transport behavior, a systematic analysis of the bias-dependent characteristics was performed for Mn-Gr/\textit{n}·\textit{h}-BN/Mn-Gr and V-Gr/\textit{n}·\textit{h}-BN/V-Gr MTJs. The study includes current-voltage (I-V) responses, TMR, and SIE under finite bias voltages ranging from $-0.5$~V to $+0.5$~V, as illustrated in Fig.~\ref{fig6} and Fig.~\ref{fig7}. The external bias ($V_\mathrm{b}$) was introduced by shifting the chemical potentials of the left and right electrodes to $+V_\mathrm{b}/2$ and $-V_\mathrm{b}/2$, respectively, thereby establishing the electrostatic potential gradient required for charge transport. 

The bias-dependent \textit{I}-\textit{V} characteristics of Mn-Gr/\textit{n}·\textit{h}-BN/Mn-Gr MTJs under P and AP magnetic configurations are presented in Figs.~\ref{fig6}(a), 6(e), 6(i), 6(m) and Figs.~\ref{fig6}(b), 6(f), 6(j), 6(n), respectively. For the monolayer \textit{h}-BN barrier [Fig.~\ref{fig6}(a)], the spin-up current in the P state increases with bias voltage and reaches a pronounced peak at $\pm0.2$~V, followed by attenuation and partial recovery beyond $\pm0.4$~V. In contrast, spin-down current remains negligible across the entire bias range, confirming highly efficient spin filtering accompanied by a pronounced negative differential resistance (NDR) effect. This NDR effect, where the current decreases with increasing bias, has significant technological relevance for high-frequency oscillators, nonvolatile memory, and ultrafast electronic switches.
In the AP configuration [Fig.~\ref{fig6}(b)], both spin channels are nearly suppressed within the bias window of $-0.2$~V to $+0.2$~V. Beyond this range, selective enhancement of one spin channel and suppression of the other emerge, indicative of spin-dependent tunneling asymmetry under AP alignment. For the trilayer \textit{h}-BN barrier ($n = 3$), the spin-up current in the P configuration exhibits an approximately linear dependence on the applied bias voltage, while the spin-down component behaves similarly to the $n = 1$ case. In the AP configuration, the current is nearly completely suppressed across the entire bias range.
 In contrast, MTJs with even-layer barriers ($n = 2$ and $n = 4$) exhibit qualitatively different characteristics. When $n = 2$, the P state current exhibits oscillatory modulation across the entire bias range. For $n = 4$, the spin-up current evolves gradually under positive bias but exhibits sharp enhancement beyond $-0.4$~V in the negative bias region.
These barrier layer-number-dependent transport properties highlight the critical role of \textit{h}-BN thickness in designing the transport functionality of vdW MTJs.

The \textit{I}-\textit{V} characteristics of V-Gr/\textit{n}·\textit{h}-BN/V-Gr MTJs under P and AP magnetic configurations are depicted in Figs.~\ref{fig7}(a), 7(e), 7(i), 7(m) and Figs.~\ref{fig7}(b), 7(f), 7(j), 7(n), respectively. A pronounced odd--even layer-number dependence emerges in the \textit{I}-\textit{V} response of few-layer \textit{h}-BN barriers, accompanied by robust spin filtering and prominent NDR effects. In the AP configuration, the system exhibits strong bias-controlled spin selectivity: spin-up current dominates under negative bias, whereas spin-down current prevails under positive bias. This voltage-gated spin-polarized transport enables dynamic control over spin-resolved conduction pathways, rendering these MTJs as high-performance, electrically programmable spin valves.
The SIE can be tuned to approach 100\% through bias voltage modulation, demonstrating precise electrical control over spin-polarized currents. As the bias increases, enhanced tunneling probability further improves spin injection while preserving spin selectivity within specific voltage windows.
The TMR behavior of Mn-Gr/\textit{n}·\textit{h}-BN/Mn-Gr MTJs is summarized in Figs.~\ref{fig6}(d), 6(h), 6(l), and 6(p). The TMR exhibits a gradual decline with increasing bias, peaking at equilibrium ($V_{\mathrm{b}} = 0$). Notably, a giant TMR of $4.35 \times 10^8$\% is achieved for the monolayer barrier ($n = 1$). 
A clear odd--even oscillation in TMR as a function of barrier thickness is observed: odd-layer barriers ($n = 1$, $3$) retain ultrahigh TMR values ($4.35 \times 10^8$\%, $2.76 \times 10^7$\%), whereas even-layer configurations ($n = 2$, $4$) show reduction by several orders of magnitude ($1.67 \times 10^6$\%, $1.12 \times 10^6$\%).
The strong layer-number dependence and voltage-tunable spin polarization collectively position these vdW MTJs as promising platforms for next-generation spintronic applications.

Similar layer-number-dependent oscillatory behavior in TMR is also observed in V-Gr/\textit{n}·\textit{h}-BN/V-Gr MTJs, as shown in Figs.~\ref{fig7}(d), 7(h), 7(l), and 7(p). Odd-layer barriers ($n = 1$, $3$) retain elevated TMR ratios of $1.75 \times 10^4$\% and $1.86 \times 10^5$\%, respectively, whereas even-layer configurations exhibit significantly reduced values (e.g., $3.61 \times 10^4$\% for $n = 2$). Intriguingly, the $n = 2$ case exhibits a nonmonotonic TMR profile, with a peak value of $4.74 \times 10^4$\% emerging under a positive bias of $V_{\mathrm{b}} = 0.5$~V, suggesting an anomalous bias-induced enhancement unique to this configuration.
These results collectively indicate that the spin
polarization and TMR behavior in both Mn- and V-intercalated MTJs are governed by the \textit{h}-BN barrier thickness. The observed layer-correlated oscillations and ultrahigh TMR magnitudes
underscore the technological potential of these vdW MTJs for advanced spintronic applications demanding
exceptional spin selectivity and voltage-tunable functionality.

\section{Conclusion}

In summary, based on first-principles calculations, we have theoretically investigated the thickness-dependent spin transport properties of Mn-Gr/$n$·$h$-BN/Mn-Gr and V-Gr/$n$·$h$-BN/V-Gr MTJs engineered via atomic intercalation. The influence of external bias and in-plane biaxial strain on spin-polarized transport was systematically explored. Remarkably high TMR ratios of $4.35\times10^8\%$ and $1.86\times10^5\%$ were achieved in Mn- and V-intercalated systems, respectively. In addition, nearly perfect spin filtering and pronounced NDR effects were observed.
These results demonstrate that atomic intercalation enables precise control over interlayer magnetic coupling and electronic band structures, offering a scalable and tunable platform for the design of high-performance 2D spintronic devices. The proposed strategy paves the way for the development of next-generation atomic-scale spintronic architectures, including ultrahigh-density magnetic memory and ultralow-power logic circuits, thereby advancing the practical application of vdW heterostructures in spintronics.

\section{Experimental Section}
All structural relaxations were carried out using the Vienna \emph{Ab initio} Simulation Package (VASP)~\cite{ref39}, within the framework of density functional theory (DFT). The exchange-correlation potential was treated using the Perdew-Burke-Ernzerhof (PBE) functional under the generalized gradient approximation (GGA)~\cite{ref40,ref41}. A plane-wave energy cutoff of 500~eV was employed. Brillouin zone sampling was performed using a $9\times9\times1$ Monkhorst-Pack \emph{k}-point mesh, and orbital occupancies were treated with a Gaussian smearing of 0.01~eV~\cite{ref42}. A vacuum spacing of 20~\AA\ was introduced along the out-of-plane direction to eliminate spurious interactions between periodic images. The structures were optimized until the total energy and atomic forces converged below $10^{-5}$~eV and $0.01$~eV/\AA, respectively.

The spin-polarized quantum transport calculations were performed using the Nanodcal package, which combines the non-equilibrium Green’s function (NEGF) formalism with DFT~\cite{ref43,ref44}. The PBE functional within the GGA was used to describe exchange-correlation interactions. A double-$\zeta$ polarized atomic orbital basis set was adopted, with a kinetic energy cutoff of 80~Hartree and a Hamiltonian matrix convergence threshold of $10^{-5}$~eV. The electronic temperature was set to 300~K via the Fermi-Dirac distribution. Transmission spectra were computed using a $500\times500\times1$ \emph{k}-mesh, and current-voltage characteristics were evaluated over a $100\times100\times1$ grid. The electrode self-consistent field calculations employed a $9\times9\times100$ \emph{k}-point sampling. Biaxial strain was applied by uniformly adjusting the in-plane lattice constants.

The spin-resolved current \( I_{\sigma} \) and the corresponding conductance \( G_{\sigma} \) were computed using the Landauer–Büttiker formalism~\cite{ref45, ref46}: 
\begin{equation}
    I_\sigma = \frac{e}{h} \int T_\sigma(E) [f_L(E) - f_R(E)] dE,
\end{equation}
\begin{equation}
G_{\sigma} = \frac{e^2}{h} T_{\sigma},
\end{equation}
where \( \sigma \) denotes the spin index (\( \uparrow \) or \( \downarrow \)). In this framework, \( e \) is the elementary charge, \( h \) is Planck’s constant, \( T_{\sigma}(E) \) represents the energy-dependent transmission coefficient for spin \( \sigma \), and \( f_{\text{L(R)}}(E) \) denotes the Fermi–Dirac distribution function of the left (right) electrode. The total current \( I \) is obtained by summing the spin-up and spin-down contributions. The spin injection efficiency (SIE) is defined as:
\begin{equation}
    \text{SIE} = \left|\frac{I_\uparrow - I_\downarrow}{I_\uparrow + I_\downarrow}\right|.
\end{equation}
The TMR in equilibrium is given by \cite{ref47}:
\begin{equation}
\text{TMR} = \frac{G_{\text{P}} - G_{\text{AP}}}{G_{\text{AP}}} = \frac{T_{\text{P}} - T_{\text{AP}}}{T_{\text{AP}}},
\end{equation}
and under a bias \( V \), it is expressed as:
\begin{equation}
\text{TMR}_{(V)} = \frac{I_{\text{P}} - I_{\text{AP}}}{I_{\text{AP}}},
\end{equation}
with \( T_{\text{P/AP}} \) and \( I_{\text{P/AP}} \) being the transmission coefficients and currents in the parallel (P) and antiparallel (AP) magnetic configurations, respectively.

\end{justify}
\medskip
\textbf{Supporting Information} \par 
Supporting Information is available from the Wiley Online Library or from the author.

\medskip
\textbf{Acknowledgements} \par 
This work was supported by the National Natural Science Foundation of China (No. 12304148), the National Natural Science Foundation of China Regional Innovation and Development Joint Fund Key Program (No. U24A6002), and the Shanxi Natural Science Basic Research Program (No. 202203021222219).

\medskip

%



\begin{figure}[htb!]
\centering
  \includegraphics[width=9cm]{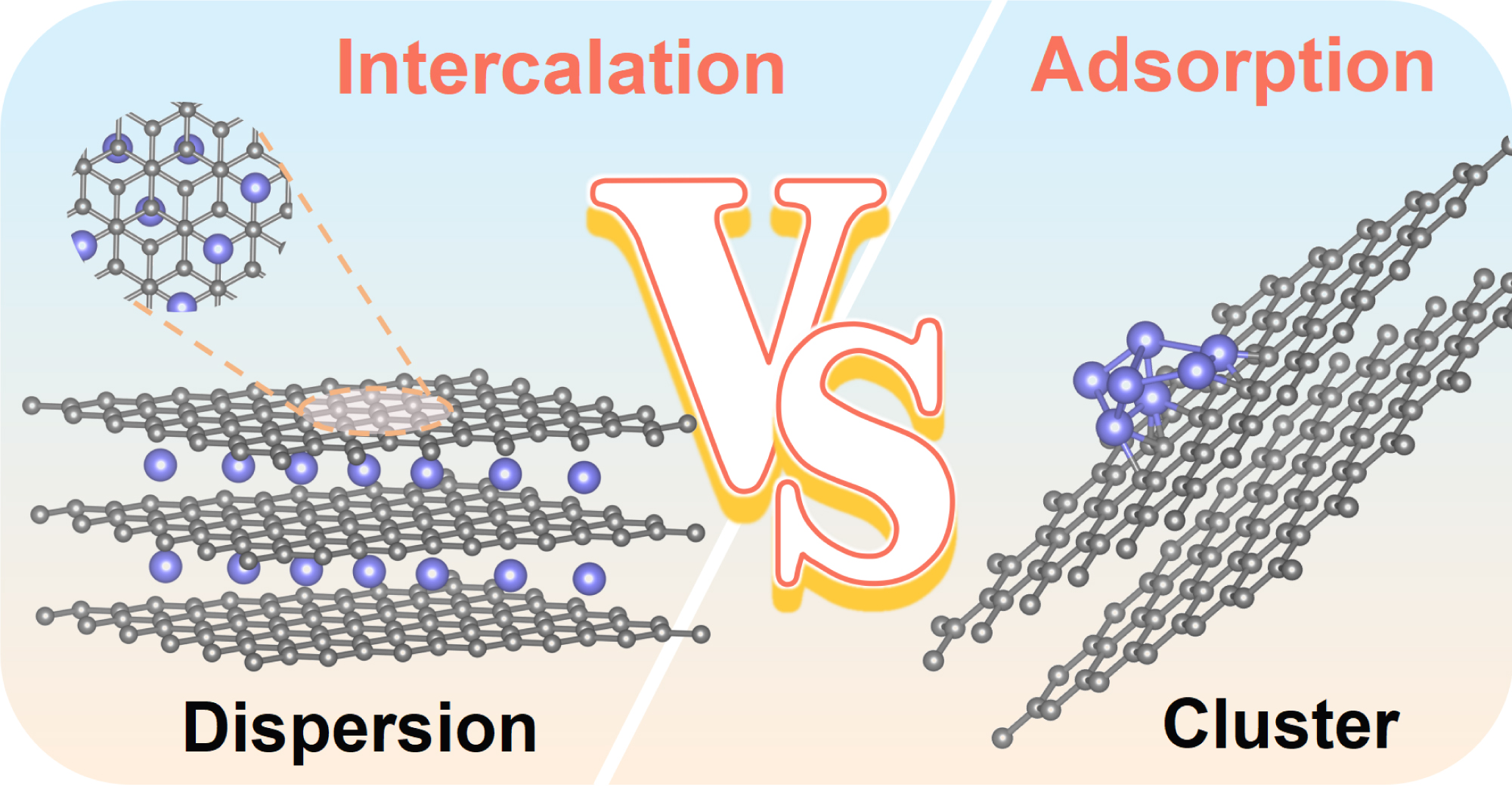}
  \caption{Comparative schematic of intercalation and adsorption mechanisms in graphene.}
  \label{fig1}
\end{figure}

\begin{figure}
  \includegraphics[width=\linewidth]{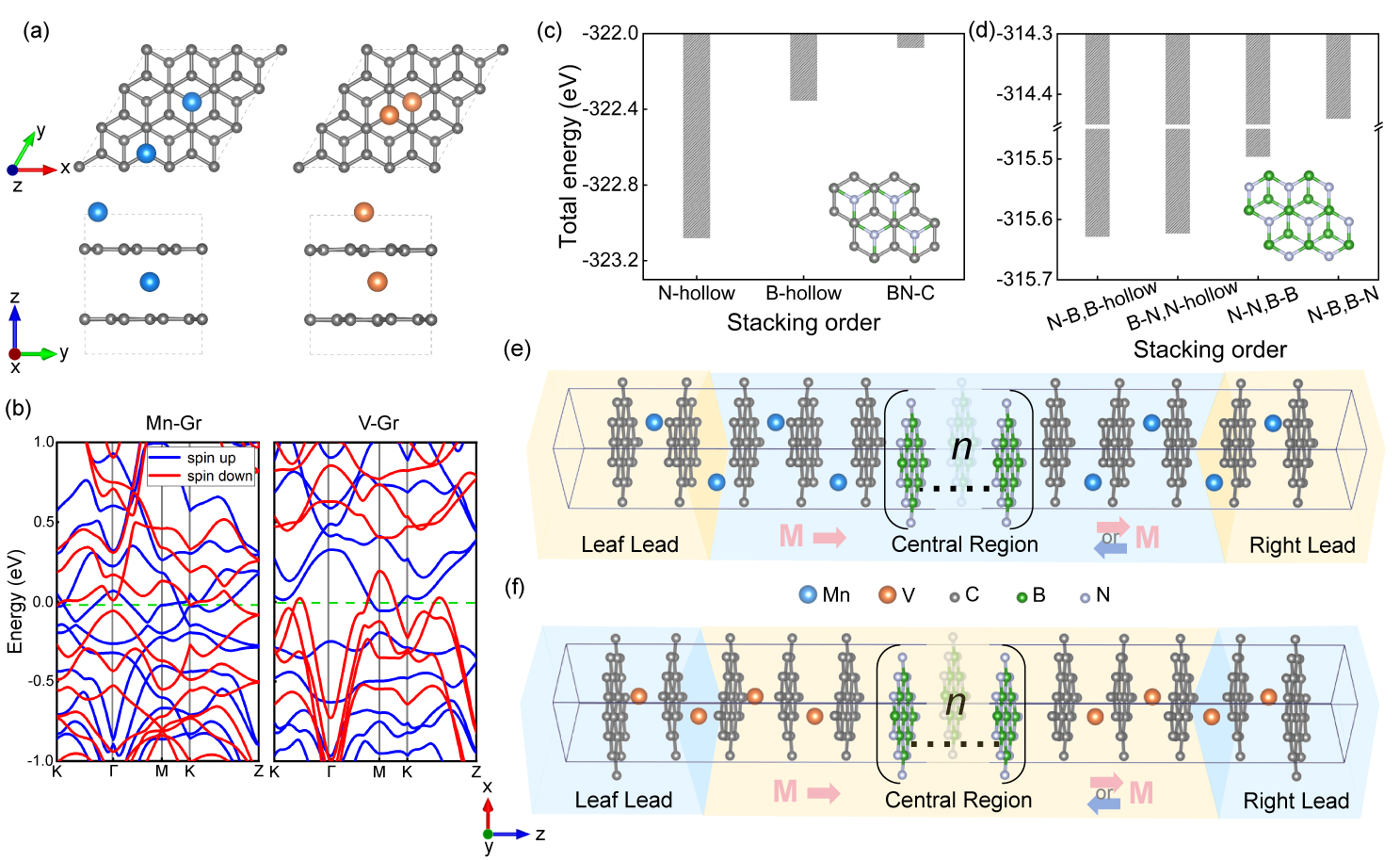}
  \caption{(a) Top and side views of the Mn-Gr and V-Gr structures. Carbon, manganese, and vanadium atoms are represented by light gray, blue, and orange spheres, respectively. 
(b) Electronic band structures of Mn-Gr and V-Gr, with the Fermi level set to 0~eV. 
(c)-(d) Total energy profiles of the Gr/$h$-BN and $h$-BN/$h$-BN interfaces under various stacking configurations. 
(e)-(f) Schematic diagrams of Mn-Gr/$n \cdot$$h$-BN/Mn-Gr and V-Gr/$n \cdot$$h$-BN/V-Gr ($n$=1-4) magnetic tunnel junctions (MTJs), respectively. 
The left and right electrodes extend to $\mp\infty$. The systems are periodic in the $xy$-plane, with current flowing along the $z$-direction.}
  \label{fig2}
\end{figure}

\begin{figure}
\centering
  \includegraphics[width=13cm]{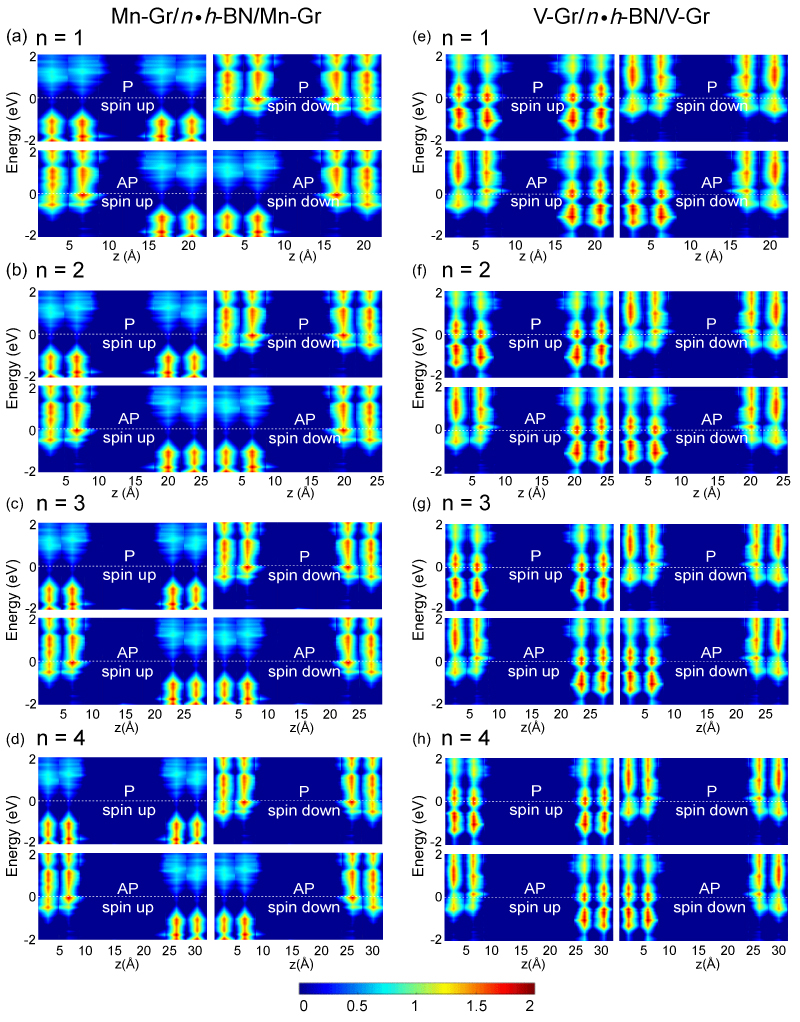}
  \caption{Spin-resolved projected density of states (PDOS) along the transport direction ($z$-axis) for the central scattering regions of Mn-Gr/$n \cdot$$h$-BN/Mn-Gr and V-Gr/$n \cdot$$h$-BN/V-Gr van der Waals magnetic tunnel junctions (vdW-MTJs) in equilibrium, with $n$=1-4. 
The Fermi level ($E_\mathrm{F}$) is indicated by white dashed lines.}
  \label{fig3}
\end{figure}

\begin{figure}
  \includegraphics[width=\linewidth]{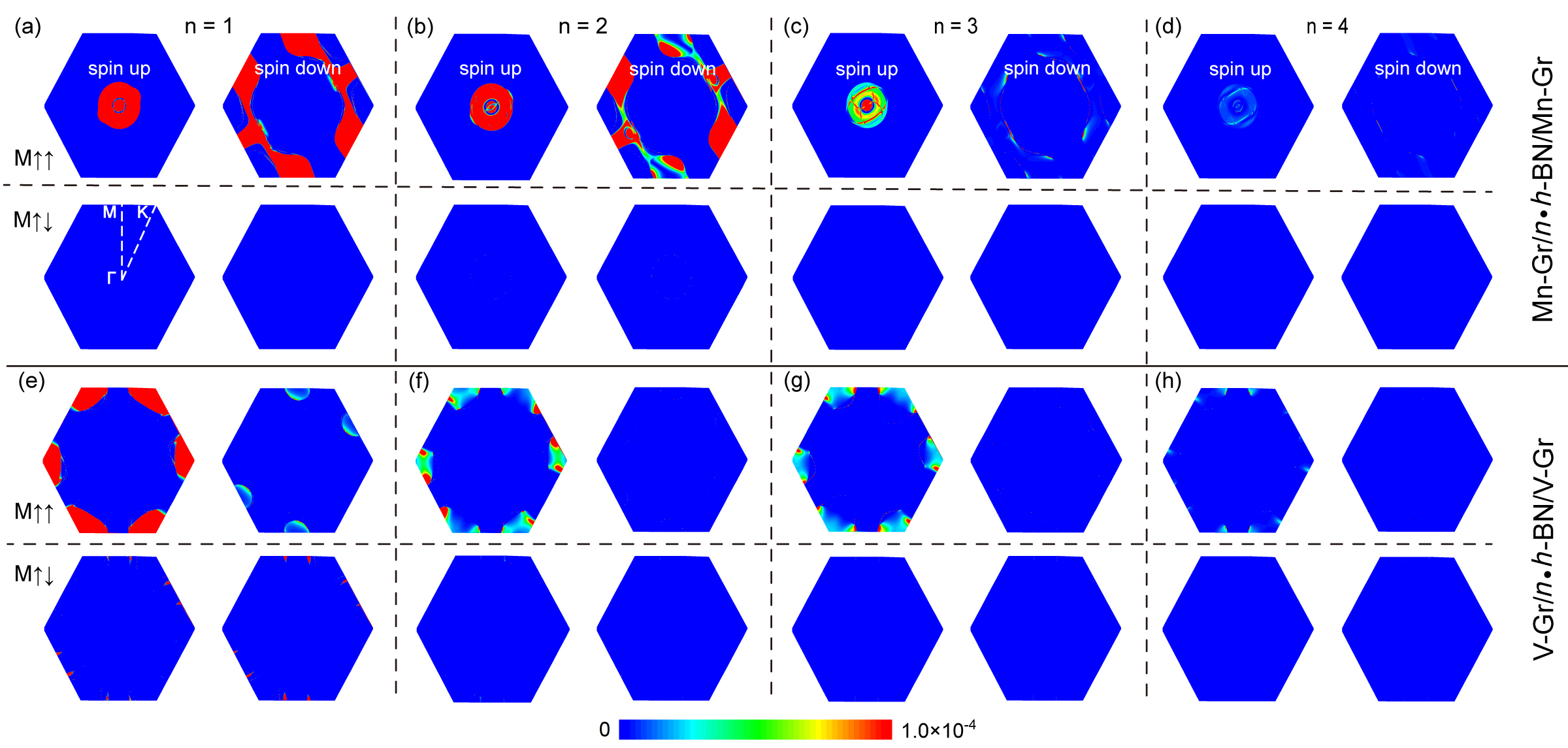}
  \caption{The $k_\Arrowvert$-resolved transmission coefficients of Mn-Gr/$n \cdot$$h$-BN/Mn-Gr and V-Gr/$n \cdot$$h$-BN/V-Gr ($n$=1-4) in the 2D Brillouin zone at the Fermi level.}
  \label{fig4}
\end{figure}

\begin{figure}
\centering
  \includegraphics[width=15cm]{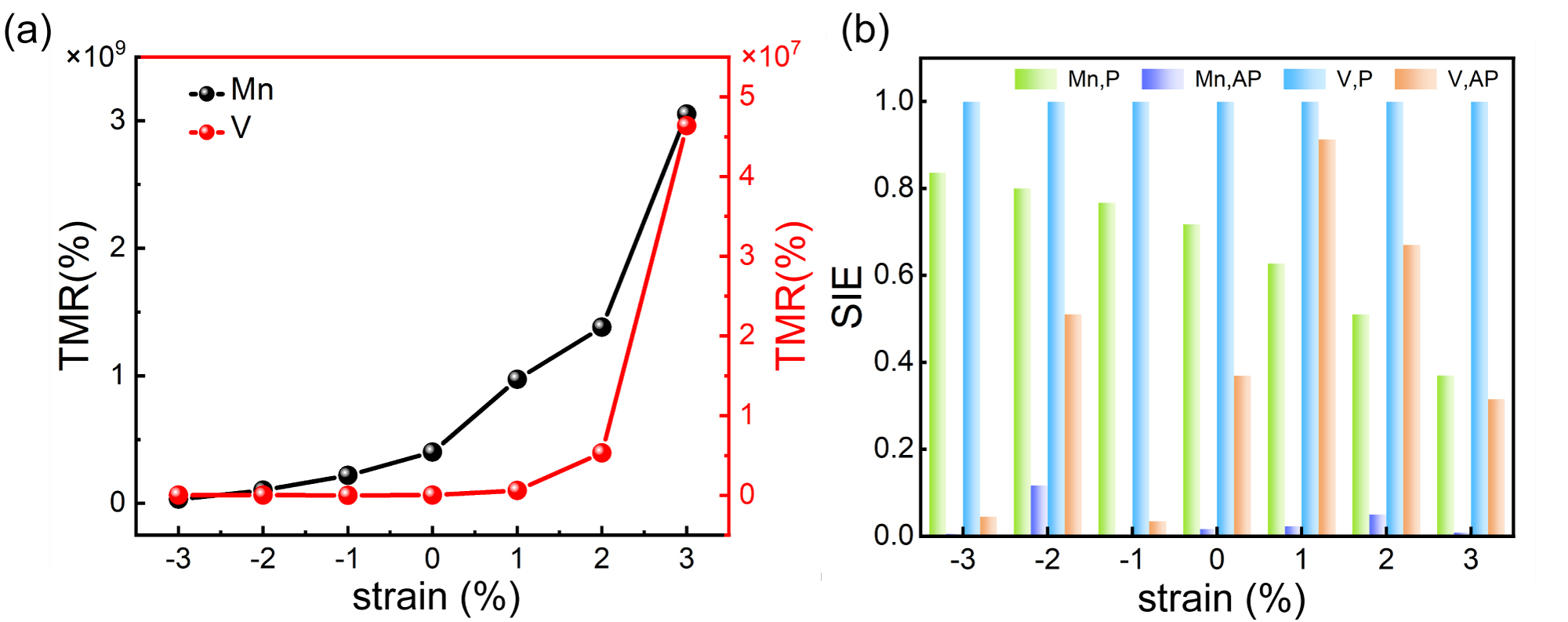}
  \caption{(a) TMR and (b) SIE of Mn-Gr/$h$-BN/Mn-Gr and V-Gr/$h$-BN/V-Gr MTJs under equilibrium conditions, plotted as functions of in-plane biaxial strain ($\varepsilon = -3\%$ to $+3\%$).}
  \label{fig5}
\end{figure}

\begin{table}[htbp]
\centering
\caption{Calculated spin-dependent electron transmission ($T_\uparrow$, $T_\downarrow$), tunnel magnetoresistance (TMR), and spin injection efficiency (SIE) for Mn-Gr/$n\cdot$$h$-BN/Mn-Gr and V-Gr/$n\cdot$$h$-BN/V-Gr magnetic tunnel junctions.}
\label{tab1}
\sisetup{
  exponent-product = \times,
  table-number-alignment = center,
  table-format = 1.2e2
}
\footnotesize 
\setlength{\tabcolsep}{4.5pt} 
\renewcommand{\arraystretch}{1.3} 
\begin{tabular}{@{} 
    >{\RaggedRight}p{1.5cm}
    c
    S[table-format=1.2e2]
    S[table-format=1.2e2]
    S[table-format=1.2e2]
    S[table-format=2.1]
    S[table-format=1.2e2]
    S[table-format=1.2e2]
    S[table-format=1.2e2]
    S[table-format=2.1]
    S[table-format=1.2e2]
    @{}}
\toprule
\multirow{2}{*}{System} & 
\multirow{2}{*}{$n$} & 
\multicolumn{4}{c}{P State} & 
\multicolumn{4}{c}{AP State} & 
\multicolumn{1}{c}{TMR (\%)} \\ 
\cmidrule(lr){3-6} \cmidrule(lr){7-10}
& & 
{$T_\uparrow$} & 
{$T_\downarrow$} & 
{$T_\text{total}$=$T_\uparrow$+$T_\downarrow$} & 
{SIE (\%)} &  
{$T_\uparrow$} & 
{$T_\downarrow$} & 
{$T_\text{total}$=$T_\uparrow$+$T_\downarrow$} & 
{SIE (\%)} \\ 
\midrule
\multirow{4}{*}{\makecell{Mn-Gr/\\$n\cdot$$h$-BN/\\Mn-Gr}} 
& 1 & 4.62e2  & 7.8e3   & 5.39e2  & 71.1 & 6.45e9  & 5.93e9  & 1.24e8  & 4.2  & 4.35e8 \\
& 2 & 9.54e3  & 1.41e4  & 2.36e4  & 19.3 & 7.00e9  & 7.08e9  & 1.41e8  & 0.6  & 1.67e6 \\
& 3 & 1.94e3  & 3.55e3  & 5.48e3  & 29.3 & 1.55e10 & 4.42e11 & 1.98e10 & 55.5 & 2.76e7 \\
& 4 & 5.91e7  & 2.12e7  & 2.71e6  & 56.5 & 2.27e10 & 1.50e11 & 2.42e10 & 87.6 & 1.12e6 \\
\midrule
\multirow{4}{*}{\makecell{V-Gr/\\$n\cdot$$h$-BN/\\V-Gr}} 
& 1 & 5.20e3  & 1.19e6  & 5.20e3  & 99.9 & 6.29e6  & 2.33e5  & 2.96e5  & 57.4 & 1.75e4 \\
& 2 & 9.16e6  & 4.46e7  & 1.01e5  & 91.1 & 4.49e7  & 3.07e9  & 4.52e7  & 98.6 & 2.12e3 \\
& 3 & 3.15e3  & 2.51e6  & 3.39e5  & 85.2 & 9.62e9  & 8.69e9  & 1.83e8  & 5.1  & 1.86e5 \\
& 4 & 6.58e7  & 1.2e8   & 6.70e7  & 96.4 & 1.50e10 & 1.70e9  & 1.84e9  & 83.8 & 3.61e4 \\
\bottomrule
\end{tabular}
\smallskip
\end{table}

\begin{figure}
  \includegraphics[width=\linewidth]{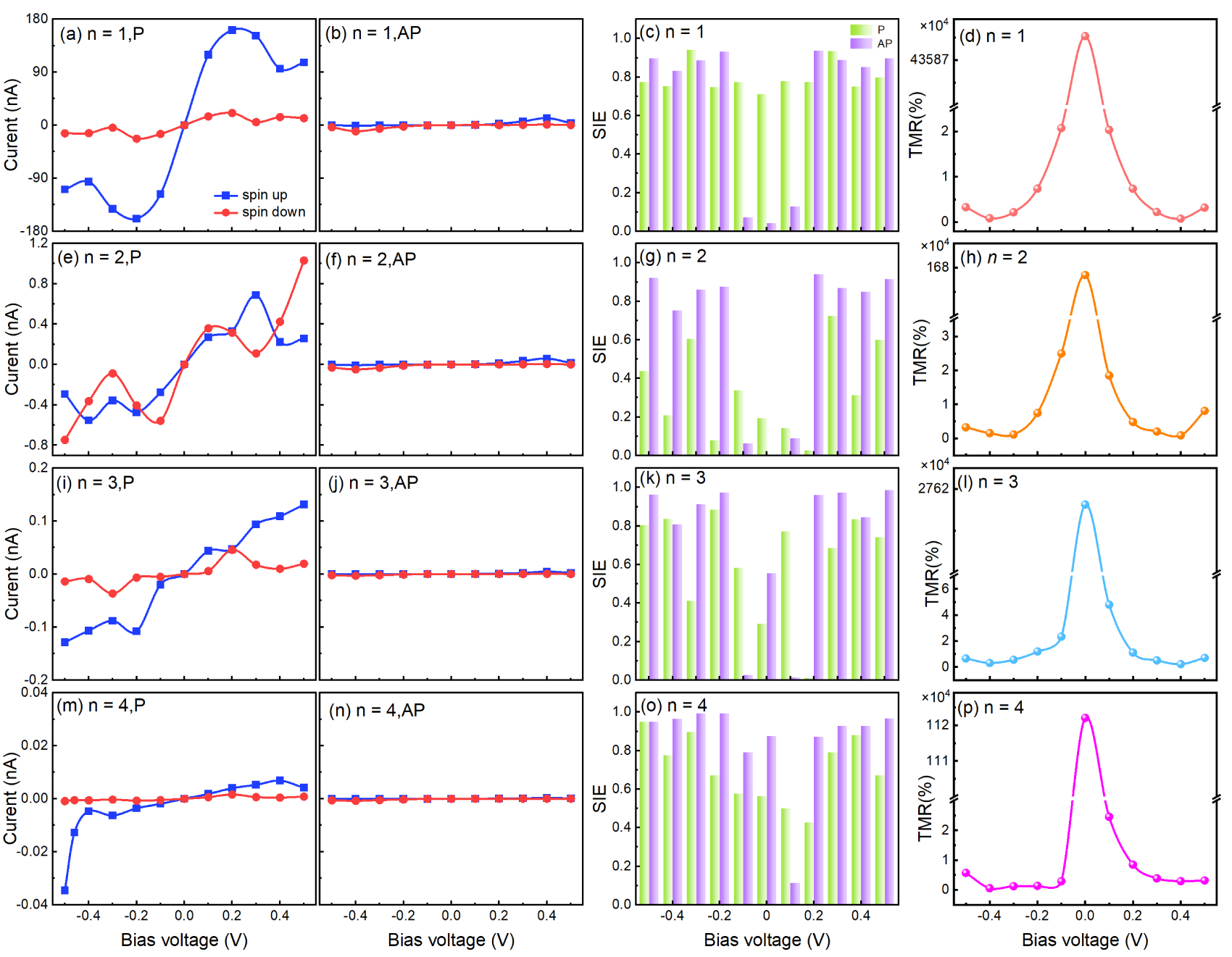}
  \caption{Transport characteristics of the Mn-Gr/$n \cdot$$h$-BN/Mn-Gr MTJs under non-equilibrium conditions. Panels (a), (b), (e), (f), (i), (j), (m), and (n) show the current as a function of bias voltage. Panels (c), (g), (k), and (o) present the SIE, while panels (d), (h), (l), and (p) display the TMR as a function of bias voltage.}
  \label{fig6}
\end{figure}

\begin{figure}
  \includegraphics[width=\linewidth]{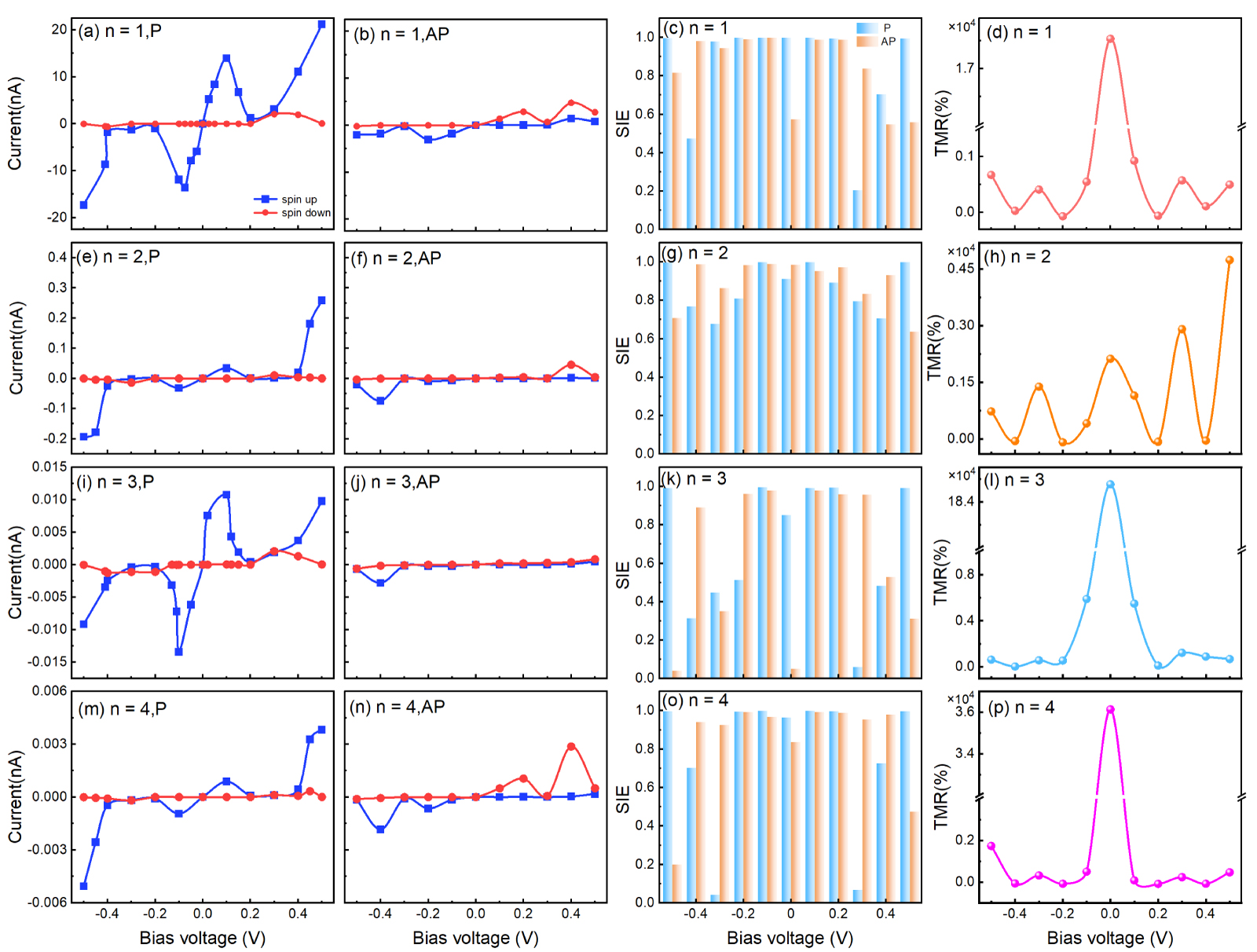}
  \caption{Transport characteristics of the V-Gr/$n \cdot$$h$-BN/V-Gr MTJs under non-equilibrium conditions. Panels (a), (b), (e), (f), (i), (j), (m), and (n) show the current as a function of bias voltage. Panels (c), (g), (k), and (o) present the SIE, while panels (d), (h), (l), and (p) display the TMR as a function of bias voltage.}
  \label{fig7}
\end{figure}



\begin{figure}
\textbf{Table of Contents}\\
\medskip
\centering
  \includegraphics[width=0.7\linewidth]{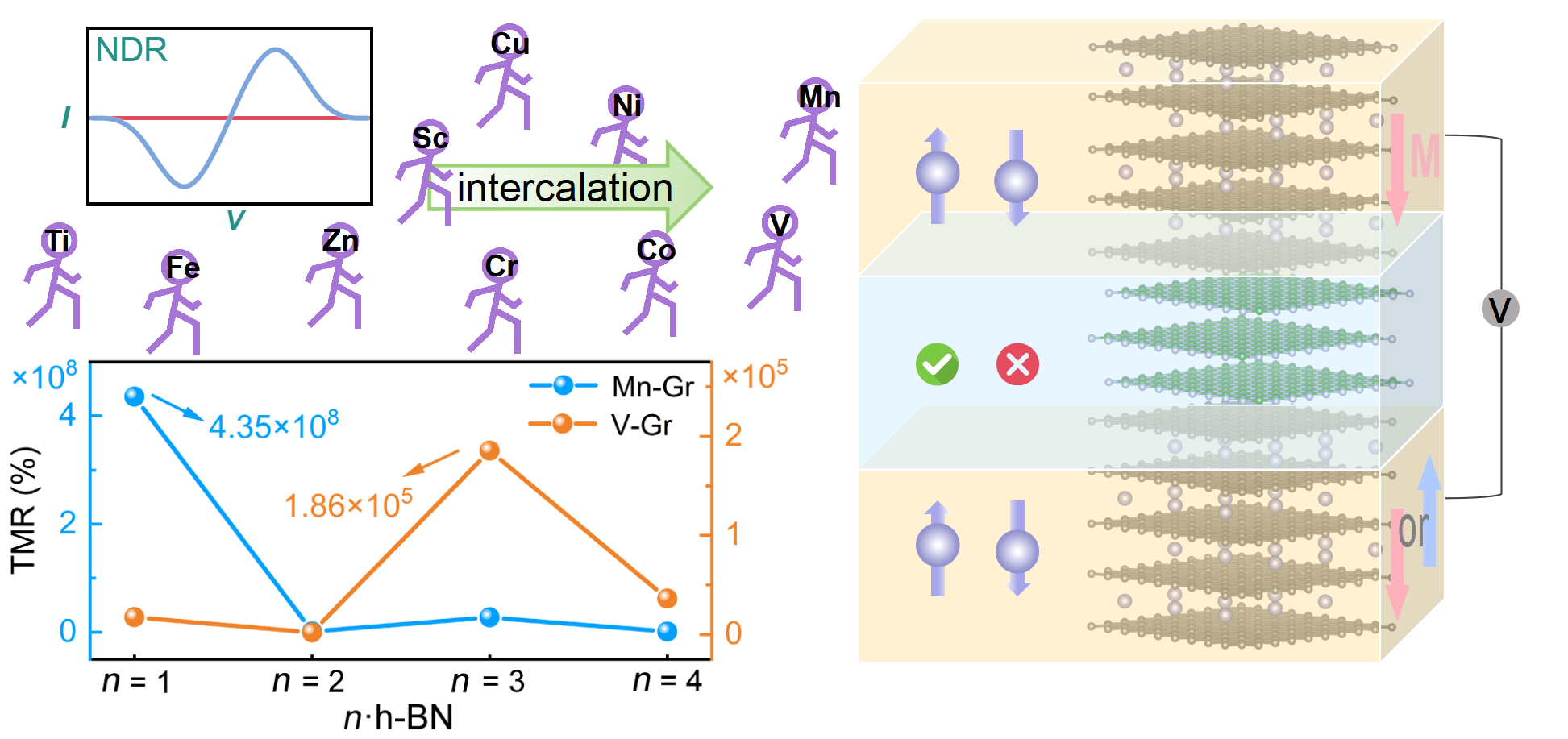}
  \medskip
  \caption*{Fully 2D magnetic tunnel junctions are designed using transition metal–intercalated graphene and $h$-BN barriers. Intercalation induces robust ferromagnetism and atomic-level dispersion. These junctions exhibit giant TMR up to $10^9\%$, perfect spin filtering, and strong negative differential resistance, enabling high-performance spintronic applications.}
\end{figure}

\end{document}